\documentclass[aps,a4paper, 11pt, onecolumn,nofootinbib]{revtex4}
 \pdfoutput=1
\usepackage{bm}
\usepackage{mathrsfs}
\usepackage{xcolor,color,graphicx,graphics}
\usepackage[all]{xy}
\usepackage{epsfig,subfigure}
\usepackage{latexsym,amssymb,amsmath,amsfonts} 
\usepackage[english]{babel} 
\usepackage[OT1]{fontenc}
\usepackage[latin1]{inputenc}
\usepackage{makeidx}
\usepackage{hyperref}
\usepackage{color,graphicx,graphics,wrapfig,epsf}

\definecolor{red}{rgb}{1,0,0}

\def\p{\partial}

\def\+{^\dagger}

\def\<{\leftarrow}
\def\>{\rightarrow}

\def\({\left(}
\def\){\right)}

\def\a{\alpha} \def\b{\beta} \def\g{\gamma} \def\d{\delta} \def\e{\epsilon}
\def\m{\mu} \def\n{\nu} \def\r{\rho} \def\s{\sigma} \def\l{\lambda} \def\th{\theta}
\def\k{\kappa}\def\D{\Delta}\def\G{\Gamma}\def\O{\Omega}\def\L{\Lambda}

\def\q{\quad}\def\qq{\qquad}

\def\E{{\cal E}}

\newcommand{\LL}{{\cal L}}

\newcommand{\bi}{\begin{itemize}} 				\newcommand{\ei}{\end{itemize}}
\newcommand{\benu}{\begin{enumerate}} 		\newcommand{\enu}{\end{enumerate}}
\newcommand{\bd}{\begin{dinglist}{0}}     \newcommand{\ed}{\end{dinglist}}
\newcommand{\bfig}{\begin{figure}[htbp]}  \newcommand{\efig}{\end{figure}}
        			
\newcommand{\bc}{\begin{center}} 				  \newcommand{\ec}{\end{center}}
\newcommand{\be}{\begin{equation}} 				\newcommand{\ee}{\end{equation}}
\newcommand{\bsub}{\begin{subequations}}  \newcommand{\esub}{\end{subequations}}
\newcommand{\ben}{\begin{eqnarray}} 			\newcommand{\een}{\end{eqnarray}}
\newcommand{\ba}[1]{\begin{array}{#1}} 		\newcommand{\ea}{\end{array}}
\newcommand{\bea}{\begin{equation}\begin{array}{rcl}}
\newcommand{\eea}{\end{array}\end{equation}}

  \makeatletter
\newcommand\xleftrightarrow[2][]{%
  \ext@arrow 9999{\longleftrightarrowfill@}{#1}{#2}}
\newcommand\longleftrightarrowfill@{%
  \arrowfill@\leftarrow\relbar\rightarrow}
\makeatother

\begin{document}
\title{Compact scalar field solutions in EiBI gravity}

\author{Victor I. Afonso} \email{viafonso@df.ufcg.edu.br}
\affiliation{Unidade Acad\^{e}mica de F\'isica, Universidade Federal de Campina
Grande, 58429-900 Campina Grande, PB, Brazil}

\date{\today}
\begin{abstract}
We discuss exact scalar field solutions describing gravitating compact objects
in the Eddington-inspired Born-Infeld gravity (EiBI), 
a member of the class of (metric-affine formulated) Ricci-based gravity theories (RBGs).
We include a detailed account of the RBGs/GR correspondence exploited to analytically solve the field equations.
The single parameter $\e$ of the EiBI model defines two branches for the solution.
The $\e>0$ branch may be described as a `shell with no interior', and constitutes an ill-defined, geodesically incomplete spacetime.
The more interesting $\e<0$ branch admits the interpretation of a `wormhole membrane', 
an exotic horizonless compact object with the ability to transfer particles and light 
from any point on its surface (located slightly below the would-be Schwarzschild radius) to its antipodal point, in a vanishing fraction of proper time.
This is a single example illustrating how the structural modifications introduced by the metric-affine formulation 
may lead to significant departures from GR even at astrophysically relevant scales, giving rise to physically plausible objects radically different from those we are used to think of in the metric approach, and that could act as a black hole mimickers whose shadows might present distinguishable signals.
\end{abstract}


\maketitle


\section{Introduction}\label{sec:I}
Besides the longstanding application to inflation\cite{ArmendarizPicon:1999rj} and accelerating solutions\cite{ArmendarizPicon:2000ah},
since the pioneering work of Fisher \cite{Fisher}, scalar field configurations have been widely explored in the context of gravitating systems.
Even though in the Einstein's General Relativity context the uniqueness theorems and the no-hair conjecture \cite{NH} state that the only three quantities describing any black hole solution (BH) would be mass, charge, and angular momentum, efforts were devoted to the search for `hairy' black holes solutions. 
These are extensions of the Kerr-Newman solution in the sense that,
besides the basic properties, they count with the extra ingredient of a `cloud' of scalar matter surrounding
 the black hole, supported by different kinds of self-interactions -- see for instance Refs. \cite{KNhair}, and also 
 \cite{HBS} and \cite{VolkovReview} for updated reviews on this topic.
 
In recent years, the interest in scalar field supported compact gravitating solutions has risen, 
fueled by several unexpected results with plausible astrophysical applications.
Rotating black holes \cite{Herdeiro:2018daq}, black hole shadows \cite{Cunha:2018acu}, 
and new black hole solutions with non-trivial scalar hair\cite{Herdeiro:2015waa}, are under investigation. 
Also, in addition to hairy black holes, other developments address the matter of scalar field supported `solitonic' configurations, such as bosons stars\cite{Colpi:1986ye, Liebling:2012fv, Palenzuela:2017kcg,  Macedo:2013jja, Cunha:2017wao},  Proca stars\cite{Brito:2015pxa}, gravitating skyrmions \cite{GS} and other topological solutions\cite{Bazeia:2007df}, gravastars\cite{Pani:2010em, Visser:2003ge, Chirenti:2016hzd}, 
and other quasi-stationary configurations around black holes \cite{NSG}, as well as other types of scalar solutions like those discussed in Refs. \cite{Ayon-Beato:2015eca,Sol-other} and \cite{GravSol}. 
Within this context, new phenomena such as `superradiance' \cite{Superradiance}, or `black hole bomb' instabilities
\cite{BBB,Cardoso:2004nk,Hod:2016kpm}  can arise, triggered by such scalar fields configurations.

Modifications/extensions of Einstein's General Relativity are motivated by many different reasons, arising from
astrophysical\cite{Barack:2018yly,SeyYag18} (Dark Matter) to cosmological\cite{Ishak:2018his} (Dark Energy) origins
---some broad reviews on the subject can be found in \cite{DeFelice:2010aj, CLreview, Clifton:2011jh, Nojiri:2017ncd, Heisenberg:2018vsk, BeltranJimenez:2019tjy}--, as well as from purely theoretical/technical issues (information loss \cite{Unruh:2017uaw, Ellis:2019}, singularities \cite{Senovilla:2006db} --see however \cite{Olmo:2015bya,Olmo:2016hey}--, etc.). 
The classical tests of GR cover the range from laboratory to Solar System scales, sensing gravity only in its weak field regime \cite{Will:2014kxa}. 
Astrophysical observations, from pulsars to stellar orbits around the galactic centers, sense gravity on a stronger regime and already put important constraints on the modified theories, that must behave much as GR at short scales.  \cite{EzZuma} 
Extragalactic \cite{Collett:2018gpf} and cosmological \cite{Ishak:2018his} contexts allow for other precision tests as well.

The recent observations by the LIGO/Virgo Collaboration of binary black hole and neutron star mergers \cite{Abbott:2016blz} inaugurated the era of gravitational waves astronomy, which allows for testing gravity in the strong regime as well as at large scales.
The data accumulation after successive runs of the detectors, along with results of near future 
galaxy surveys such as EUCLID \cite{Laureijs:2011gra,Amendola:2016saw} or  LSST \cite{lsst},
 will probe GR well beyond the classical tests and will pose more stringent limits on the viable extensions of Einstein's theory.
Actually, the binary neutron star merger recently detected by its multi-messenger (gravitational plus electromagnetic) signal
has already ruled out or put hard constraints on several popular extensions \cite{Lombriser:2015sxa, Lombriser:2016yzn, Baker:2017hug,Sakstein:2017xjx,Creminelli:2017sry}.
Now, the existence of compact objects of more exotic kinds is a possibility that cannot be disregarded at the present early stage of the GWs astronomy.\cite{Cardoso:2019rvt}
In fact, it seems to be the right moment to look for new non-canonical solutions and investigate their physical implications,
as they could act as markers guiding the identification of deviations from standard GR objects, that may arise in future astrophysical observations.\cite{Barack:2018yly,SeyYag18,fs}
In particular, the existence of compact objects without a horizon whose signals could be degenerated with that of black holes (Black Hole Mimickers), and how they could be discriminated through observations is a question that is under current intense investigation \cite{Cardoso:2017cqb}. 

Gravitating compact scalar field solutions have already been investigated in the context of modified gravity\cite{Carballo-Rubio:2018jzw}. 
Moreover, if dark matter is described by axions (or ultra-light bosons), clouds of this matter could be formed around black holes \cite{Arvanitaki:2016qwi}. 
However, the aforementioned new set of observational capabilities points to the need for a thorough review of the principles on which the modifications of GR shall be based on, and has given rise to a number of different proposals.
One such alternative approach is the metric-affine or `Palatini' formulation of theories of gravity, in which metric and affine connection are considered as {\it a priori} independent and equally fundamental objects to be determined dynamically\cite{Olmo:2011uz}.

In spite of having very appealing features like, for instance, second order field equations for any theory,
the metric-affine approach has been relatively much less explored as compared to the metric extensions of GR.
The main obstacle in the development of these kind of theories resides, on the one hand, in the highly non-linear character of the field equations which one has to deal with. On the other hand, obtaining direct explicit solutions for the connection equation could result in an absolutely daunting task.
These features have limited the exploration to the use of specific {\it ans\"atze} or approximative methods.
Besides, despite the long and fruitful development of Numerical Relativity, which heavily contributed to the success of the field in recent years, the coding has evolved based on the \emph{metric} structure of the Einstein's equations 
for which the ($3+1$) Hamiltonian formalism and BSSN formulation were developed.
This makes the direct implementation of very well tested numerical tools to problems of physical interest in the metric-affine approach, unfeasible in terms of computational and programming costs (see, for instance,  Ref. \cite{Berti:2018vdi} for a discussion on the case of gravitational wave emission of binary mergers). 

After all those difficulties, hope was partially recovered when it was realized that, at least for models 
involving only the Ricci tensor and the metric (Ricci-based Gravities or RBGs), part of the connection equation can be solved by the introduction an `auxiliary metric'. \cite{Olmo:2011uz}
Thus, metric-affine extensions of GR have been kept mainly within the bounds of this class of models--see, however,  \cite{Burton:1997sj} and \cite{Deruelle,Borunda:2008kf,Charmousis08}, where scalar-tensor models incorporating nonmetricity, and Lovelock theories are considered, or the recent study of Horndeski Lagrangians in \cite{Helpin:2019kcq}.
Fortunately, the universe of RBGs theories is wide enough to make it worth its exploration since, amid other extensions, it contains any $f(R)$-type model, the Eddington inspired Born-Infeld family of `squared-root' theories\cite{Banados:2010ix} and, of course, the Palatini version of GR itself. 
\footnote{About the longstanding belief on the complete equivalence between the metric and metric-affine treatment of Einstein-Hilbert action in vaccum, supported by Einstein itself \cite{FF82}, see however \cite{Buchdahl} and the stunning recent results in \cite{Bejarano:2019zco}.}

Further work in that direction lead us to the identification of an Einstein frame description for the RBGs class of models\cite{Afonso:2018bpv}.
Remarkably enough, the transformation leading to the new frame introduces no extra degrees of freedom but, instead, 
new relationships between the geometric and the matter components are unveiled.
In this sense, this transformation may be 
better interpreted as a duality relation linking distinct theories.
This map relates the solution space of arbitrary RBG metric-affine models, including their matter fields,
to the space of solutions of GR coupled to the same kind of matter but driven by different dynamics.
Remarkably, in each and every studied case, the underlying non-linear structure of the RBG gravity Lagrangian density
ends up transferred by the map into the matter field sector, giving rise to non standard matter Lagrangians coupled to GR.

A similar behavior is observed in the inverse map.
In fact, the strength of this technique resides in that 
all the relations are \emph{purely algebraic} and, being established at the level of the field equations, they're not restricted to any particular symmetry or solution.
Also, at least formally, the relation must be invertible. Hence, whenever an exact solution of one of the theories is known, 
it allows for the obtention of exact solutions for the related theory. 
Even more, in practical terms, it unlocks all the known numerical techniques and facilities available to deal with GR problems
to be exploited in favor of modified gravity theories formulated in the metric-affine approach.
The mechanism is now well understood and, after being successfully implemented in different systems including general anisotropic fluids \cite{Afonso:2018bpv}, electromagnetic fields \cite{Afonso:2018mxn}, and scalar fields \cite{Afonso:2018hyj}, has shown to be very robust.
Also, its remarkably efficiency is well exemplified when comparing the awkward direct resolution 
of the field equations of an RBG plus scalar matter system presented in \cite{Afonso:2017aci}, 
to the elegant and straightforward re-derivation using the mapping procedure shown in \cite{Afonso:2018bpv}, also discussed below.

\medskip
The main aim of the present work is 
to discuss gravitating compact scalar field solutions, obtained by exploiting the Einstein frame representation of the Eddington-inspired Born-Infeld gravity.
Interestingly, some of these objects could be interpreted as black hole mimickers.
In order to get there, we will go through the details of the mapping algorithm, using as a seed to feed the mechanism the well known spherically symmetric solution obtained by Wyman in \cite{Wyman}, which corresponds to a static configuration of a single real scalar field
described by a canonical Lagrangian density coupled to GR.

\medskip The manuscript is organized as follows.
In Sec. \ref{sec:II} we delimit properly what we call the class of Ricci-based Gravity theories (RBGs), and derive their field equations on general grounds;
in Sec. \ref{sec:III} we build the Einstein-frame representation, applying the mapping prescription to a scalar field coupled to an arbitrary Ricci-based gravity, which we later particularize to the EiBI gravity case in Sec. \ref{sec:IV}.
Then, after briefly reviewing the Wyman's solution of GR, 
exact solutions  for the EiBI model  are obtained and their features analyzed in Sec. \ref{sec:V}.
Finally, Sec.\ref{sec:VI} is reserved to summarize our results and discuss some future perspectives of this line of research.

\section{Ricci-based gravities} \label{sec:II}
Consider the class of theories described by actions of the form
\be \label{eq:actionRBG}
\mathcal{S}=
\int d^4x  \sqrt{-g} \mathcal{L}_{G}\(g_{\mu\nu}, R_{(\mu\nu)}(\Gamma)\) 
+\int d^4x \sqrt{-g} \mathcal{L}_m(g_{\mu\nu},\psi_m) \ .
\ee
The gravity sector is described by a Lagrangian density  $\mathcal{L}_{G}$ given in terms of 
scalars constructed upon the metric  $g_{\mu\nu}$ and the Ricci tensor $R_{\m\n}(\G)\equiv\p_{\a}\G^{\a}_{\n\m}-\p_{\n}\G^{\a}_{\a\m}+\G^{\a}_{\a\b}\G^{\b}_{ \n\m} -\G^{\a}_{\n\b}\G^{\b}_{\a\m}$, which depends only on the affine connection, here assumed to be an independent variable (metric-affine formalism).
Actually, we will only consider the symmetric part of the Ricci tensor, $R_{(\m\n)}$, \footnote{Parenthesis will be removed from now on for notational simplicty.}  
to guarantee that the theory is free of the ghost-like instabilities that may arise
when the antisymmetric part of the Ricci tensor takes place in the action.\cite{BeltranJimenez:2019acz}
The matter sector is described by the Lagrangian density $\mathcal{L}_m(g_{\mu\nu},\psi_m)$, with $\psi_m$ denoting collectively the matter fields, which corresponds to minimally coupled bosonic fields. 
In fact, there is no direct coupling between the matter fields and the connection.
This choice preserves the Equivalence Principle, in the sense that matter particles will follow geodesics determined by the metric alone. But besides, this is so in order to preserve the projective invariance of the theory, which allows to remove the torsion by a gauge choice and, thus, it can be safely disregarded from the onset (for details on this subtle point we remit the reader to Ref.\cite{Afonso:2017bxr}).

The kind of theories under the above prescription is what we call Ricci-based gravity theories (RBGs).
As mentioned, the pool of RBGs is wide enough to contain the Einstein-Hilbert action, any $f(R)$, $f(R,R_{\mu\nu}R^{\mu\nu})$, among others the like, but also more sophisticated constructs as the EiBI gravity, that we will deal with in this work.

\subsection{Field equations}
Let us now study carefully the field equations of this class of models.
The Palatini variation (metric and connection varied independently) of action \eqref{eq:actionRBG} leads to the two tensorial equations
\ben
2 \frac{\p \LL_G}{\p g^{\m\n}}- {\LL_G} g_{\m\n} = T_{\m\n}  \label{eq:gmn}\\
 \nabla^\G_\a \left(\sqrt{-g}  \frac{\p \LL_G}{\p R_{\m\n}}\right)=0  \ . \label{eq:dGamma}
\een
The RHS of equation \eqref{eq:gmn},  the one related to the metric variation,
gets the usual stress-energy tensor of the matter, $T_{\m\n}\equiv\frac{-2}{\sqrt{-g}}\frac{\d (\sqrt{-g}\LL_m)}{\d g^{\m\n}}$, while
the affine connection related equation \eqref{eq:dGamma} is homogeneous, as a result of our choice of minimally coupled matter fields.

Differently from the metric formulation, where the relation between the metric and the  affine connection (Christoffel's symbols) is an {\it a priori} prescription, here we need to work out a solution of \eqref{eq:dGamma}. Depending on the actual form of $\LL_G$, this can be a virtually impossible task.
However, a clever definition can shift the problem to a more convenient configuration that had allowed for a resolution in every and all the cases considered until now.
The move consist in introducing a new object whose relation to the connection is already known through the form of its equation, namely,
to define the tensor
\be\label{eq:defq}
\sqrt{-q}\, q^{\m\n}\equiv 2\k^2 \sqrt{-g}  \frac{\p \LL_G}{\p R_{\m\n}}\ ,
\ee
where $q\equiv\det{q_{\m\n}}$ (under the implicit assumption that $q\neq0$, and thus $q^{\m\n}$ has an inverse, $q_{\m\n}$), and $\k^2$ is Newton's constant in adequate units (in GR we have $\kappa^2=8\pi G$).
Then, equation \eqref{eq:dGamma} reads $ \nabla^\G_\a \left(\sqrt{-q} q^{\m\n}\right)=0 $, which makes the covariant derivative compatible with the `auxiliary metric' $q_{\m\n}$. That is, $\G$ is of Levi-Civita type, a connection whose components are the Christoffel's symbols of $q_{\m\n}$, namely, $ \G_{\m\n}^{\l}=\frac{1}{2} q^{\l\a}(\p_{\m}q_{\n\a}+\p_{\n}q_{\m\a}-\p_{\a}g_{\m\n})$.

We have thus formally solved the connection equation in terms of $q_{\m\n}$, but we still need to understand how the affine connection relates to the physical metric $g_{\m\n}$. To that aim note that $q_{\m\n}$ is symmetric by definition, as long as $\LL_G$ is built on the symmetrized Ricci tensor. 
Let us then propose the existence of an \emph{algebraic} relation between $q_{\m\n}$ and the space-time metric $g_{\m\n}$, 
realized by a matrix $\hat\O$ such that
\be \label{eq:Omegadef}
q_{\m\n}=g_{\m\a}{\O^\a}_{\n} \ .
\ee
Therefore, $\sqrt{-q}={|\hat\O|}^{\frac12}\sqrt{-g}$, with $\vert\hat\Omega\vert\equiv\det{\hat\Omega}$, and also $(\O^{-1})^{\m}_{\ \n}=q^{\m\a}g_{\a\n}$.

To see how this definition plays, note that, being $\LL_G$ a scalar density, it can only be made out of powers of traces the object $g^ {\m\b}R_{\b\n}(\G)$ and, therefore, derivatives of $\LL_G$ with respect to both, $g_{\m\n}$ and $\G^{\a}_{\m\n}$, will present a common factor. 
Besides, in virtue of \eqref{eq:defq}, on shell $R_{\a\b}(\G)=R_{\a\b}(q)$.
One can thus trace Eq. \eqref{eq:gmn} with $g^{\m\n}$, and use relation \eqref{eq:Omegadef} 
to express any instance of $g_{\m\n}$ on its RHS in terms of $q_{\m\n}$ and ${\O^\a}_{\n}$.
This allows to write the metric field equation in the very appealing form
\be\label{eq:RBGeom}
G^{\m}_{\ \nu}(q)={\k^2}{\vert \hat\O \vert^{-\frac12}} \left[ T^{\m}_{\ \n} -\(\tfrac{T}{2}+\LL_G  \) \d^{\m}_{\ \n} \right] \ ,
\ee
where ${G^\mu}_{\nu}(q)\equiv q^{\m\a}(R_{\a\n}(q)-\frac12 q_{\a\n}R(q))$, is the Einstein tensor of the auxiliary metric $q_{\m\n}$,
while $T^{\m}_{\ \n}= g^{\m\a} T_{\a\n}$ and $T$ is its trace.

\smallskip 
It is worth recalling here that Eqs. (\ref{eq:RBGeom}) are of second-order by construction,
enjoying a ghost-free character while recovering exactly the GR form in vacuum (${T^\mu}_{\nu}=0$).
This is an extremely important point, as it is precisely this feature what prevents the RGB class of theories
of propagating any extra degrees of freedom beyond those of GR (gravitational waves with two polarizations traveling at the speed of light).\footnote{The fact that RBGs gravitational waves propagate along the null geodesics of $q_{\mu\nu}$, 
and its impact on the tensorial stability of regular solutions in high-energy density environments was analyzed in \cite{BeltranJimenez:2017uwv,Bazeia:2015zpa}. Also, RBG GWs propagation in a FRW background and possible bounds on the model internal parameters was discussed in \cite{Jana:2017ost}.}

\section{Einstein frame}\label{sec:III}
Despite the appealing form of the field equations \eqref{eq:RBGeom},
a serious drawback of this kind os theories is that it is not always possible to
go from the formal relation \eqref{eq:Omegadef} to an 
explicit expression for $g_{\mu\nu}$ in terms of the auxiliary metric  $q_{\mu\nu}$.
 This makes very difficult to extract information from physically relevant systems, besides specific highly symmetric cases as, for instance
\cite{Olmo:2011ja,Olmo:2013gqa,Bambi:2015zch} in astrophysical, or \cite{Odintsov:2014yaa,BeltranJimenez:2017uwv}, in cosmological contexts.
However, the form in which we have expressed relation \eqref{eq:RBGeom} has the clue to advance towards the solution of the conundrum.
On one hand, as will be explicitly shown soon, once a particular RBG Lagrangian $\LL_G$ is chosen, 
the matrix $\hat\Omega$ can be written as a function of the matter fields and the metric $g_{\m\n}$,
and this \emph{on-shell} property is also shared by $\mathcal{L}_{G}$ itself.
Even when the nonlinearities of relation \eqref{eq:RBGeom} are much worst than in any GR system
(while Einstein's equations are linear in $T^\m_{\ \n}$
here the RHS shows an intricate dependence on the matter fields),
our equations \eqref{eq:RBGeom} show the `correct'  `geometry = matter' form present in the GR equations, to which they exactly reduce when $T^\m_{\ \n}=0$.

These facts motivated the proposal presented in \cite{Afonso:2018bpv},
that can be explained as a reinterpretation of the RHS of \eqref{eq:RBGeom} as the 
{\it actual} stress-energy tensor of a different matter model coupled to an Einstein-Hilbert action.
 That is, a GR system described in terms of the metric $q_{\m\n}$ in proper Einstein frame.
Namely, we propose to read Eq. \eqref{eq:RBGeom} as $G^{\m}_{\ \nu}(q)=\k^2 \tilde T^{\m}_{\ \nu}$, where 
\be\label{eq:RBGeom2}
\tilde T^{\m}_{\ \nu}\equiv {\vert \hat\O \vert^{-1/2}} \left[ T^{\m}_{\ \n} -\(\tfrac{T}{2}+\LL_G  \) \d^{\m}_{\ \n} \right] \ .
\ee
Naturally, this identification can only be valid under the 
necessary condition that $\tilde T^{\m}_{\ \nu}= q^{\m\a} \tilde T_{\a\nu}$ define a \emph{conserved} stress-energy tensor in the new frame,
which is also a consistency requirement of the contracted Bianchi identities, $\nabla^{\G}_{\mu} \,{G^{\mu}}_{\nu}(q)=0$.
That condition is perfectly fulfilled by minimally-coupled 
matter sources such as generic anisotropic fluids\cite{Afonso:2018bpv},
matter Lagrangians constructed out of one or several scalar fields\cite{Afonso:2018hyj}, or electromagnetic fields\cite{Afonso:2018mxn}.

The above relation can be stated as follows:
for any given RBG model $\LL_G$, (minimally) coupled to a matter Lagrangian $\LL_m$ with stress-energy tensor $T_{\m\n}$,
it is possible to find another Lagrangian $\tilde \LL_m$ with stress-energy tensor $\tilde T_{\m\n}$, coupled to GR, whose solution space is in full correspondence with that of the RBG theory. And, moreover, the converse is also valid.
That is, departing from GR coupled to some matter field, its solutions can be related to the ones of an RBG theory coupled to another matter field of the same kind, but described by a different (in general non-linear) Lagrangian -- see Table \ref{tab:mapscheme} for an schematic representation.
\begin{table}[h]  \label{tab:mapscheme} 
\caption{Schematics of the mapping relations.}{\begin{tabular}{ccc} \toprule
 RBG& &GR\\
\colrule\\
$\int d^ 4x\sqrt{-g}\,\LL_{G}+\int d^ 4x\sqrt{-g}\,\LL(X,\phi)$
&&$\int d^ 4x\sqrt{-q}\, R+\int d^ 4x\sqrt{-q}\, \tilde \LL(Z,\phi)$ \\\\
$G^\m_{\ \n}(q)={\k^2}{\vert\O\vert^{-\frac12}} \left[ T^\m_{\ \n} -(\LL_{G}+\frac{T}{2})\right]$
&&$G^\m_{\ \n}(q)=\k^2 \tilde T^\m_{\ \n}$\\\\
$T^\m_{\ \n}=\frac{-2}{\sqrt{-g}} \frac{\p(\sqrt{-g} L(X,\phi))}{\p g_{\m\n}}$
&&$\tilde T^\m_{\ \n}=\frac{-2}{\sqrt{-q}} \frac{\p(\sqrt{-q} \tilde L(Z,\phi))}{\p q_{\m\n}}$ \\[2.4mm]
&$q_{\m\n}=g_{\m\a}\O^{\a}_{\ \n}$&\\
 $\qq\left.\vert\O\vert^{-\frac12}\! \left[ T^\m_{\ \n} -(\LL_G+\frac{T}{2})\right]\right\vert_{g_{\m\n} \>\,q_{\m\n}}$
&$\xleftrightarrow{\qq\qq\qq\q}$ & $\tilde T^\m_{\ \n}$  \\[3mm]
\botrule
\end{tabular}}
\end{table}

A remarkable property of the interrelation between these theories, unveiled by the mapping procedure,
is that the non-linear structure present in the RBG gravity Lagrangian gets mapped to 
a similar nonlinear realization, but in the \emph{matter} sector.
This manifests in a particularly striking way when considering determinantal actions, 
as the Eddington inspired Born-Infeld gravity we will consider below, 
where the square-root structure ends up related to an analogous square-root form in the matter Lagrangian. 
This feature may have a direct relevant application in discussing non-canonical scalar fields models--see \emph{e.g.} \cite{ArmendarizPicon:1999rj,ArmendarizPicon:2000ah,Bazeia:2007df}--through the different optic of metric-affine theories.

\subsection{Mapping RBGs with a scalar field}
As it is the focus of the present work, let us now explicitly construct the mapping in the case of scalar matter constituted by a single real scalar field.
We start by considering a generic (non-canonical) action of the form
\be \label{eq:scalarRBG}
\mathcal{S}_m(X,\phi)=-\frac{1}{2}\int d^4x\sqrt{-g}L(X,\phi) \ ,
\ee
where $L$ is an arbitrary function of the scalar field and the quadratic kinetic term $X=g^{\a\b}\p_\a\phi\p_\b\phi$.
The corresponding stress-energy tensor reads
\be\label{eq:TmnX}
T^{\m}_{\ \n}= L_X X^{\m}_{\ \n}- \tfrac12 {L(X,\phi)}\d^{\m}_{\ \n}\ ,
\ee
where $L_X \equiv dL/dX$,  $X^{\m}_{\ \n}\equiv g^{\mu\alpha}\partial_\alpha\phi\partial_\nu\phi$
and, therefore, $X=X^{\a}_{\ \a}$ is its trace. 

This is a very convenient form of expressing the stress-energy tensor since
beneath the central assumptions of the mapping prescription it is the possibility of writing ${\Omega^\mu}_\nu$, the matrix connecting the spacetime metric $g_{\mu\nu}$ and the Einstein frame representation metric $q_{\mu\nu}$, as a (nonlinear) function of ${T^\mu}_\nu$. 
To that aim we write an {ansatz} for ${\Omega^\mu}_\nu$ in the form of as a series expansion on ${T^\mu}_\nu$, which formally reads
\be
\hat\O=\sum c_n(X,\phi) \,\hat T^{n}\ . 
\ee
Now, a key point here is that the object $X^{\m}_{\ \n}$ introduced in \eqref{eq:TmnX} 
shows an idempotency property, namely, $(\hat X/ X )^n=\hat X/ X$, and thus, $\hat X^n= X^ {\m}_{\ \a_1} X^ {\a_1}_{\ \a_2}\cdots X^ {\a_{n-1}}_{\ \n}=X^{n-1} \hat X$.
As a result of this, any power of the stress-energy tensor \eqref{eq:TmnX} will be given as a linear combination of 
$\d^{\m}_{\ \n}$ and $X^{\m}_{\ \n}$, which let us write
\be\label{eq:OmegaX}
{\O^\m}_\n=a(X,\phi){\d^\m}_{\n}+b(X,\phi){X^\m}_\n \ ,
\ee
 where the particular form of the functions $a$ and $b$ will depend on the model under consideration.
It is precisely this structure of the matrix relating $g$ and $q$ what guarantees that every instance of $g_{\m\n}$ in the RHS of Eq.\eqref{eq:RBGeom} can be eliminated in favor of $q_{\m\n}$. Explicitly, from definition \eqref{eq:Omegadef} and representation \eqref{eq:OmegaX}, we have that
\be
g^{\m\a}\p_\a\phi = q^{\m\b} \O^{\a}_{\ \b}\p_\a\phi = 
q^{\m\b} (a\, \d^{\a}_{\ \b}+b\, X^{\a}_{\ \b})\p_\a\phi 
=(a+b X) q^{\m\b}\p_\b\phi
\ee
Therefore, we can always write 
\be\label{eq:Xmn2Zmn}
X^{\m}_{\ \n} = (a+b X ) Z^{\m}_{\ \n}\ ,
\ee
with $Z^{\m}_{\ \n}\!\equiv\! q^{\s\m}\p_\s\phi\,\p_\n\phi$.
Tracing and inverting \eqref{eq:Xmn2Zmn} we obtain $Z=X/{(a+bX)}$, where $Z\equiv {Z^\mu}_\mu$.
The function $Z=Z(X,\phi)$, is the necessary step to obtain the inverse relation, $X=X(Z,\phi)$.
Therefore, every term in Eq.\eqref{eq:RBGeom} depending on $g_{\m\n}$ (through $\LL_{G}$, $T^{\m}_{\ \n}$ and its trace $T$) 
can be put in terms of $q_{\m\n}$.
We are now finally allowed to interpret $\tilde T_{\m\n}$ as an actual stress-energy tensor of 
matter fields in the $q$-related spacetime.
In fact, we can assume the existence of a scalar field model described by an action
\be\label{eq:scalarGR}
\tilde{\mathcal{S}}_m(Z,\phi)=-\frac{1}{2}\int d^4x\sqrt{-q} \tilde L(Z,\phi) \ ,
\ee
with an associated standard stress-energy tensor  ${\tilde{T}^\m}_{\ \ \n}=\tilde L_Z Z^{\m}_{\ \n}- \tfrac{1}{2} \tilde L(Z,\phi) {\d^\m}_\n$.
Relation \eqref{eq:RBGeom2} can then be fulfiled by solving the equations
\ben
\tilde L_Z Z^{\m}_{\ \n}&=&|\hat{\O}|^{-\frac12} L_X X^{\m}_{\ \n} \label{eq:offdiag}\\
\tilde L(Z,\phi)&=&|\hat{\O}|^{-\frac12} \left(2 \LL_G +X L_X- L(X,\phi) \right)\ , \label{eq:diag}
\een
which correspond to the matching of its diagonal and non-diagonal parts.
Note that the traces of \eqref{eq:Xmn2Zmn} and \eqref{eq:offdiag} combine to give
\be\label{eq:Kz}
\tilde L_Z= |\hat{\O}|^{-\frac12} {L_X (a+b X)}  \ .
\ee
Now, to properly establish the mapping between theories coupled to the scalar matter, 
the corresponding evolution equations of the scalar fields,
\ben
2 \p_\m\left(\sqrt{-g} L_X g^{\m\a}\p_\a\phi\right)-\sqrt{-g} {L_\phi}&=&0 \label{eq:sf1}\\
2 \p_\m\left(\sqrt{-q} \tilde L_Z q^{\m\a}\p_\a\phi\right)-\sqrt{-q} {\tilde L_\phi}&=&0 \label{eq:sf2}\ .
\een
should be consistently satisfied by the solution of \eqref{eq:offdiag}-\eqref{eq:diag}.
 Making use of the determinant of Eq. \eqref{eq:Omegadef} and 
the traces of \eqref{eq:Xmn2Zmn} and \eqref{eq:offdiag}, we verify that
\be 
\sqrt{-g}L_Xg^{\m\a}\p_\a\phi 
= \sqrt{-q}  (a+b X) \tilde L_Z q^{\m\a}\p_\a\phi = \sqrt{-q} \tilde L_Z q^{\m\a}\p_\a\phi\ ,
\ee
which merges \eqref{eq:sf1} and \eqref{eq:sf2} into the single relation 
\be\label{eq:potentials} \tilde L_\phi = |\hat{\O}|^{-\frac12} L_\phi \ , \ee
where $\tilde L_\phi\equiv\p_\phi\tilde L(Z,\phi)$.
This, along with \eqref{eq:offdiag} and \eqref{eq:diag}, completes the system of equations posed by the mapping prescription.

\smallskip In cases where we are able to explicitly invert relation \eqref{eq:Xmn2Zmn} to obtain $X=X(Z,\phi)$, 
substituting this in Eq.\eqref{eq:diag} will directly give us the Lagrangian $\tilde L(Z,\phi)$.
However, this may not always be the most practical approach to take.
Instead, we can use the fact that partial derivatives of the Lagrangian $\tilde L(Z,\phi)$ namely, $\tilde L_Z$ and $\tilde L_\phi$,
must be identical to the RHS of \eqref{eq:Kz} and  \eqref{eq:potentials}, respectively. 
Thus, we can first calculate $\tilde L_X$ directly from Eq. \eqref{eq:diag} and then compare it to the product $\tilde L_Z Z_X$, 
 where $\tilde L_Z$ is given in \eqref{eq:Kz} and $Z_X$ can be derived from the trace of \eqref{eq:Xmn2Zmn}.

The final piece to have a completely consistent framework
requires one to realize that $\tilde L_\phi=\p_\phi\tilde L(Z,\phi)$, can be expressed as
\be \partial_\phi \tilde L(Z,\phi)=\partial_\phi \tilde L(X,\phi)-\tilde L_Z  Z_\phi \ ,\ee
where $L_Z$ is given in (\ref{eq:Kz}), while $Z_\phi$ can be computed from \eqref{eq:Xmn2Zmn} (recall that $a=a(X,\phi)$ and $b=b(X,\phi)$).

\section{Eddington-inspired Born-Infeld gravity}\label{sec:IV}
We will now put into practice the machinery developed in the previous section 
focusing on a specific RBG: the Eddington-inspired Born-Infeld gravity theory, 
which has been extensively studied in the last years\cite{Vollick:2003qp,Banados,BI1,BI2,BI4,BI5,BI5b,BI6,BI7,Gu:2018lub,BI8,Bouhmadi-Lopez:2018sto} (for a recent comprehensive review see \cite{BeltranJimenez:2017doy}).
It is described by the action
\be\label{eq:actionEiBI}
\mathcal{S}_{EiBI}=\frac{1}{\e\k^2} \int d^4 x \left[\sqrt{-\vert g_{\m\n} + \e R_{\m\n}(\G)\vert}-\l \sqrt{-g}\right] \ .
\ee
Two new objects, $\e$ and $\l$, appear here. The length-squared $\e$ parameter, assumed to be small, allows to show that EiBI gravity deviates from GR solutions only at high-curvature (or high-energy density) situations. In fact, a perturbative expansion in $\e$ of the above action shows that, at curvature scales $\vert R_{\mu\nu} \vert \ll 1/\epsilon$, EiBI gravity yields GR+ $\L_{eff}+\mathcal{O}(\e)$, where  $\Lambda_{eff}=\frac{\lambda-1}{\epsilon \kappa^2}$ acts as an effective cosmological constant. This feature is what guarantees the validity of EiBI theory with respect to the recently reported\cite{BI5c} equality between the speed of gravitational and electromagnetic waves propagation in vacuum.
The role of the dimensionless parameter $\l$ is clear now and, for the present work, it will be kept fixed to $\lambda=1$,
in order to capture asymptotically flat solutions.
\subsection{EiBI+scalar matter.}
The EiBI field equations for the metric \eqref{eq:gmn} take the form
\be
\sqrt{-q} q^{\m\n}- \l\sqrt{-g}  g^{\m\n}=-\e\k^2 \sqrt{-g}\, T^{\m\n}\,,  \label{eq:metric}
\ee
where the role of the auxiliary metric is assumed by the object inside the first square root in \eqref{eq:actionEiBI}, namely,  $q_{\mu\nu}\equiv g_{\mu\nu} + \epsilon R_{(\mu\nu)}(\Gamma)$, which is symmetric by construction.

Using the relation $q_{\m\n} =  g_{\m\a}\O^{\a}_{\ \n}$ we can write from \eqref{eq:metric}  an explicit equation for the matrix $\hat\O$ as
\be
|\hat{\O}|^{\frac{1}{2}}(\O^{-1})^{\m}_{\ \n}
=\l {{\d^\m}_{\n}}-\e\k^2{T^\m}_\n \ .\label{eq:Om-EiBI}
\ee
For a scalar field matter sector  the stress-energy tensor is given by \eqref{eq:TmnX}, 
and we have
\be\label{eq:Omega-1BIX}
|\hat{\O}|^{\frac{1}{2}}(\O^{-1})^{\m}_{\ \n}= A(X,\phi)\, \d^{\m}_{\ \n}+B(X,\phi) X^{\m}_{\ \n} \ ,
\ee
with $A= \lambda+\tfrac{\epsilon\kappa^2}{2} L(X,\phi)\,$ and $\,B= -\epsilon\kappa^2 L_X$.
Calculating first the determinant\footnote{Recall that $\,|\hat{M}_{4\times 4}| =\tfrac{1}{24}[{\rm tr}(\hat{M})^4-6 tr(\hat{M}^2) {\rm tr}(\hat{M})^2+8 {\rm tr}(\hat{M}^3) {\rm tr}(\hat{M})+3 {\rm tr}(\hat{M}^2)^2-6 {\rm tr}(\hat{M}^4)]\ .$
}
 we can invert relation \eqref{eq:Omega-1BIX} to obtain
\be
{\O^{\m}}_\nu =  a(X,\phi) {\d^\m}_{\n}+b(X,\phi) {X^\m}_\n \ ,
\ee
where $a= \sqrt{A(A+BX)}\,$, $\,b= -|A|B/a\,$ and the determinant reads $|\hat{\O}|= A^2 a^2$.

\medskip\noindent{\bf From EiBI to GR.}
The above expressions are all functions of $\phi$ and $X$, that is, are related to the RBG (EiBI) Lagrangian $L(X,\phi)$.
In order to establish the map, we need to rewrite them in terms of the GR variables of the $\tilde L(Z,\phi)$ Lagrangian. 
To that aim note first that, given the definition of $\hat\O$ and the form of $q_{\m\n}$ in the present case, the EiBI Lagrangian in action \eqref{eq:actionEiBI} can be written as $\LL_G=(|\hat{\O}|^\frac12-\l)/\e\k^2$. Then, using \eqref{eq:offdiag}-\eqref{eq:diag} in combination with \eqref{eq:Omega-1BIX},
we can write
\be\label{eq:Omega-1BIY}
(\O^{-1})^{\mu}_{\ \nu}= \tilde{A}(Z,\phi)\, \d^{\m}_{\ \n}+ \tilde{B}(Z,\phi) {Z^\mu}_\nu\\
\ee
where $\tilde A=1-\tfrac{\e\k^2}{2}(\tilde L-Z \tilde L_Z)\, $ and $\,\tilde B= -\e\k^2 \tilde L_Z$.
This function can be easily inverted to obtain
\be
{\O^{\m}}_\n =  \tilde a(Z,\phi){\d^\m}_{\n}+\tilde b(Z,\phi){Z^\m}_\n  \label{eq:OmZ} 
\ee
with $\,\tilde a=\!1/{\tilde A}$, $\,\tilde b= -\tilde B/\tilde A(\tilde A+\tilde B Z)$, and 
$\,|\hat{\Omega}|=\! \tilde A^{\! -3}(\tilde A+\tilde B Z )^{-1} = \tilde a^3(\tilde a+\tilde b\, Z )$. 

\medskip\noindent{\bf Inverse problem: from GR to EiBI.}
One of the more promising applications of the mapping procedure 
comes from the possibility of inverting the process.
That is, departing from a known system of some scalar field matter $\tilde L(Z,\phi)$ coupled to GR,
to identify the dynamics derived from a distinct Lagrangian $ L(X,\phi)$ describing the same kind of matter, but coupled to an RBG.\footnote{For instance, an exotic matter Lagrangian coupled to GR can be reinterpreted as matter with canonical dynamics but in a metric-affine space} 
In the EiBI gravity case, this can be attained straightforwardly by using Eqs.\eqref{eq:diag}-\eqref{eq:offdiag} to write the EiBI Lagrangian all in terms of GR fields as
\be\label{eq:PZBIa}
L(X,\phi) = 2 \LL_{G}+|\hat{\Omega}|^{1/2} (Z \tilde L_Z-\tilde L) \ ,
\end{equation}
where $\LL_G=(|\hat{\O}|^\frac12-\l)/\e\k^2$, and $|\hat{\Omega}|$ is to be expressed using \eqref{eq:OmZ}. We find
\be\label{eq:PZGen}
L(X,\phi)=\tfrac{2}{\e\k^2} \left[
\left[1-\tfrac{\e\k^2}{2} (\tilde L-Z \tilde L_Z)\right]^{-\frac12} \!\left[1-\tfrac{\e\k^2}{2} (\tilde L+Z \tilde L_Z)\right]^{-\frac12}-\lambda
\right]\ .
\ee
To make explicit the relation between $Z$ and $X$ we need to specify the form of our matter Lagrangian.

\subsection{A simple scalar field model}
The results obtained above are consistent for generic scalar matter Lagrangians $L(X,\phi)$.
Nonetheless, let us consider a slightly more concrete case (still quite general) with the 
(non-canonical) form 
\be
L(X,\phi)=p(X)-2V(\phi) \ ,
\ee
where $p(X)$ is an arbitrary function and $V(\phi)$ a scalar potential.
Then, the corresponding Lagrangian in the Einstein frame reads
\be\label{eq:exprabove}
\tilde L(X,\phi)=\tfrac{2}{\e\k^2} \left(1-\tfrac{a+ b}{\sqrt{a b^3} }\right) \qq\text{with}\qq
Z(X,\phi)= \frac{2 X\vert a\vert}{\sqrt{a b^3}} \ ,
\ee
the parametric form based on the objects 
\be 
a=b-2 \e\k^2  X p_X \qq \text{and}\qq \,b= 2\lambda+\epsilon\kappa^2( p(X)-2 V(\phi)) \ .
\ee

\medskip
We can also illustrate the inverse problem by investigating, for instance, how the \emph{canonical} scalar field model
\be\label{eq:LZEF}
\tilde L(Z,\phi)= Z - 2 V(\phi) \ ,
\ee
coupled to GR, gets mapped into the EiBI framework. 
We simply write Eq. \eqref{eq:PZGen} using this Lagrangian,
and the determinant for EiBI gravity obtained from \eqref{eq:OmZ}
to get
\be\label{eq:PZBIb}
L(Z,\phi)=\tfrac{2}{\e\k^2}
\left[\left[(1+\epsilon \kappa ^2 V(\phi )) (1-Z \epsilon \kappa ^2+\epsilon \kappa ^2 V(\phi ))\right]^{-\frac12}-\lambda\right]\ .
\ee
The relation between the kinetic terms reads
\be\label{eq:XofZBI}
Z=\left(1+\epsilon\kappa^2 V(\phi )\right)\frac{X }{1+\epsilon\kappa^2 X} \ .
\ee
Therefore, inserting \eqref{eq:XofZBI} into \eqref{eq:PZBIb} we finally get
\be\label{eq:PZBIc}
L(X,\phi)=\tfrac{2}{\e\k^2} \left(\tfrac{\sqrt{1+ \e\k^2 X}}{1+\e\k^2 V(\phi )}-\l\right) \ .
\ee

The curious feature mentioned before, namely, the trading of non-linearities between the gravitational and the matter sectors,
is shown in its full glory in the free field case ($V(\phi)\to 0$), as \eqref{eq:PZBIc} reduces to
\be\label{eq:PZBId}
L(X)=\tfrac{2}{\e\k^2} (\sqrt{1+ \e\k^2 X}-\l )\ ,
\ee
which exactly recovers the square-root form of matter Born-Infeld-like theories\cite{BI34,Felder:2002sv,Jana:2016uvq}.
This effect was repeatedly observed in the different non-linear structures studied.

\smallskip
Before calculating the full exact solutions for the EiBI model, let us briefly analyze the asymptotical behavior
in the free canonical scalar field configuration, ($p(X)=X$ with $V=0$).
In this case it can be shown that the weak-field expansion of \eqref{eq:exprabove} reads $\tilde L(Z)\approx Z+\tfrac{\e\k^2}{4}Z^2$.
The strong-field regime is a bit more involved, and depends on the sign of the parameter $\e$. 
For $\e>0$, we find that there exists a limiting value $X_{max}=1/\e\k^2$, above which $\tilde L(Z)$ becomes complex. 
Thus, the linear approximation remains a good one over the domain ending at $Z_{Max}=(2/\sqrt{27})/\e\k^2$. 
For $\e<0$, the domain of $X$ is again bounded, this time by $X_{max}=\vert 2/\epsilon\kappa^2 \vert$. However, $Z$ is now  unbounded from above. 
Indeed, in that asymptotic limit, we find that $\tilde L(Z)\approx (\vert 2/\e\k^2 \vert +Z/2+(3/2)Z^{1/3})/ \vert \e\k^2 \vert^{2/3}$ is a very good approximation and precisely the fact that the dominant term in this regime is the linear one,
explains why the approximative solution found in \cite{Afonso:2018bpv} and the numerical results of \cite{Afonso:2017aci} were in so good agreement with the exact analytical solutions we will construct in the following section.

\section{Solutions.}\label{sec:V}
{\bf Wyman's solution.} 
In GR, a static spherically symmetric spacetime get completely determined by two independent metric functions.
In the early 80s, Wyman obtained his solution\cite{Wyman} for a massless scalar field coupled to GR in that 
symmetric case by using a clever substitution based on the simplicity of the scalar field equation, $\phi_{yy}=0$,
the solution of which can be taken to be, without loss of generality, $\phi=y$.
This allows to write the line element in the convenient form
\be
ds_{GR}^2=-e^{\nu}dt^2+{e^\nu}W^{-4}dy^2 +W^{-2}(d\theta^2+\sin\theta^2d\varphi^2)  \ ,
\ee
where $\n=\n(y)$ and $W=W(y)$ are functions of the radial coordinate $y$ which, in the asymptotically flat case, 
take the closed form
\be\label{eq:nuW}
e^\n = e^{\b y} \qq \text{and} \qq
W = \g^{-1} e^{\b y/2}\sinh(\g y) \ ,
\ee
with $\g\equiv \sqrt{\b^2+2\k^2}/2$, $\b =-2GM$, and $M$ is the asymptotic Newtonian mass of the solution.
We bring your attention to the fact that in the coordinates used by Wyman, that we adopt here, 
the center of the spherical solution is reached at $y\to\infty$ while the asymptotic limit, coincident with the region where $\phi\to0$, is achieved when $y\to0$.

The equivalent RBG problem, namely, EiBI gravity coupled to a \emph{canonical} free scalar field ($\tilde L(Z)=Z$) was studied by a direct approach, with a lot of pain and moderate success, in \cite{Afonso:2017aci}.
The mapping procedure described in the preceding section allows us to serve ourselves on the Wyman's solution to efficiently generate exact solutions for the EiBI case.

\subsection{Free scalar field in EiBI gravity.} 
In the previous section we have derived the 
inverse map of a canonical free scalar field model in GR,
leading to the non-canonical scalar field Lagrangian density in EiBI gravity of Eq. \eqref{eq:PZBId}.
Then, putting $\tilde L(Z)=Z$, the deformation matrix of Eq.\eqref{eq:Omega-1BIY} reduces to
${(\O^{-1})^{ \m}}_\n=  {\d^\m}_{\n}-\e\k^2 {Z^\m}_\n \,$
which, from \eqref{eq:Omegadef},  implies
\be\label{eq:Omega-1map}
g_{\m\n}= q_{\m\n}-\e\k^2 Z_{\m\n} \ .
\ee
Because of the scalar field solution $\phi(y)=1$, we have that $Z_{\m\n}=\p_{\m}\phi\p_{\n}\phi=\d_{\m y}\d_{\n y}$.
Consequently, the EIBI line element assumes the strikingly simple form
\be\label{eq:BImapping}
ds_{EiBI}^2=-e^{\n}dt^2+\left({e^\n}{W^{-4}}-\e\k^2\right)dy^2 
+W^{-2}(d\th^2+\sin\th^2d\varphi^2)  \ .
\ee
This is quite a surprising and, apparently, completely innocuous result:
the only modification with respect to the GR solution is a slight constant shift $\e\k^2$ in the $y$--$y$ component of $g_{\m\n}$.
A second look on this, however, quickly changes one's mind after realizing that bold physical consequences arise, as we will show next.

\subsection{Properties of the solution} 
Using \eqref{eq:nuW}, the radial function in \eqref{eq:BImapping} can be written as
\be
r^2(y)= W^{-2}(y) = \g^2\,\text{csch}^2\left(\g y\right) e^{-\b y} \,, \q\text{with}\q \g= \tfrac12 \sqrt{\b^2+2\k^2}
\ee
This function has a monotonic behavior, as it goes from $r^2(y) \simeq 1/y^2 $ in the asymptotic limit ($y \to 0$), to $r^2(y)\approx 
4\g^ 2 e^{-\left(2\g+\beta \right)y}$ by the central region ($y\to \infty$).

Now, as the shifting `direction' in \eqref{eq:BImapping} depends on the sign of the Born-Infeld parameter $\e$, there are two different branches to explore.

\subsubsection{Case $\epsilon >0$.}
The first thing to note in this case is that, while $g_{tt}=-e^{\n}$ is always negative, 
$g_{yy}={e^\n}{W^{-4}}-\e\k^2$ can become negative at \emph{finite} values of $y$.
Thus, the physically acceptable spacetime region is restricted to the domain of $y$ values for which $g_{yy}\ge 0$.
In the range of astrophysical configurations ($\vert\b\vert^2 \gg \k^2$), the critical value can be found to be
$y_c \approx -|\b|^{-1}\log\left(|\e|\k^2/{\b^4}\right)$.
This allows to calculate the radius of the solution at this critical point.
The expansion of the radial function  around this point (still in the astrophysical regime) can be written as
\be
r_c^2(y_c) \approx \beta^2 -\left(\log\left[\frac{\beta^4}{\vert \epsilon  \vert \kappa^2}\right]-2\right)\kappa^2 + \mathcal{O}(\kappa^4)\ .
\ee
Clearly, $r_c$ is always smaller than $|\b|=2M$, that is, than the Schwarzschild radius of an object with the same mass in GR.

These results seem to be bad news for the physical viability of this kind of object.
While its GR cousin can be interpreted as a `compact ball of scalar energy' distributed in the whole region with $r\lesssim 2M$ (and down to the center $r=0$), our EiBI object, having practically identical properties in the external region, is ill-defined for $r < r_c$. 
Besides, its Ricci scalar diverges at $y_c$ (polynomially as $R\sim 1/(\epsilon\kappa^2(y-y_c)^2)$), as well as the components
of the Einstein tensor, which implies unbounded effective energy density and pressures
(For the GR solution such divergences grow exponentially with $y$ as $y\to \infty$ ($r\to 0$)).
Finally, the spacetime represented by this configuration is geodesically incomplete, 
as can be readily verified by noting that any radial null geodesic can't be extended beyond the $r=rc$ surface, which is reached in a finite affine time.

\subsubsection{Case $\e<0$.}
 A quick look on the line element \eqref{eq:BImapping} in this case shows that, on one hand,
the solution is well defined all the way down to the center ($y \to \infty$), as $g_{yy}$ is nonvanishing (positive definite).
On the other hand, it quickly goes to a constant as $y\to\infty$ ($g_{yy}\to |\epsilon|\kappa^2$). 
Thus, the proper radial distance to the center is infinite, namely
\be\label{eq:properdistance}
\lim_{y\to\infty} L=\int^y \sqrt{g_{yy}}dy\sim \sqrt{|\e|\k^2}y\to \infty \ .
\ee

Differently to the previous case, here the internal region ($r<|\beta|=2M$) shows a much more gentle behavior, as the Ricci scalar curvature approaches to a constant, roughly estimated as $R(y \to \infty) \to -{\beta ^2}/{2 |\epsilon| \kappa^2 }$
for astrophysical objects.
This result is illustrated in Fig.\ref{fig:Ricci}, showing solutions with different $\b$ (mass) values.

\begin{figure}[bp]
\centerline{\psfig{file=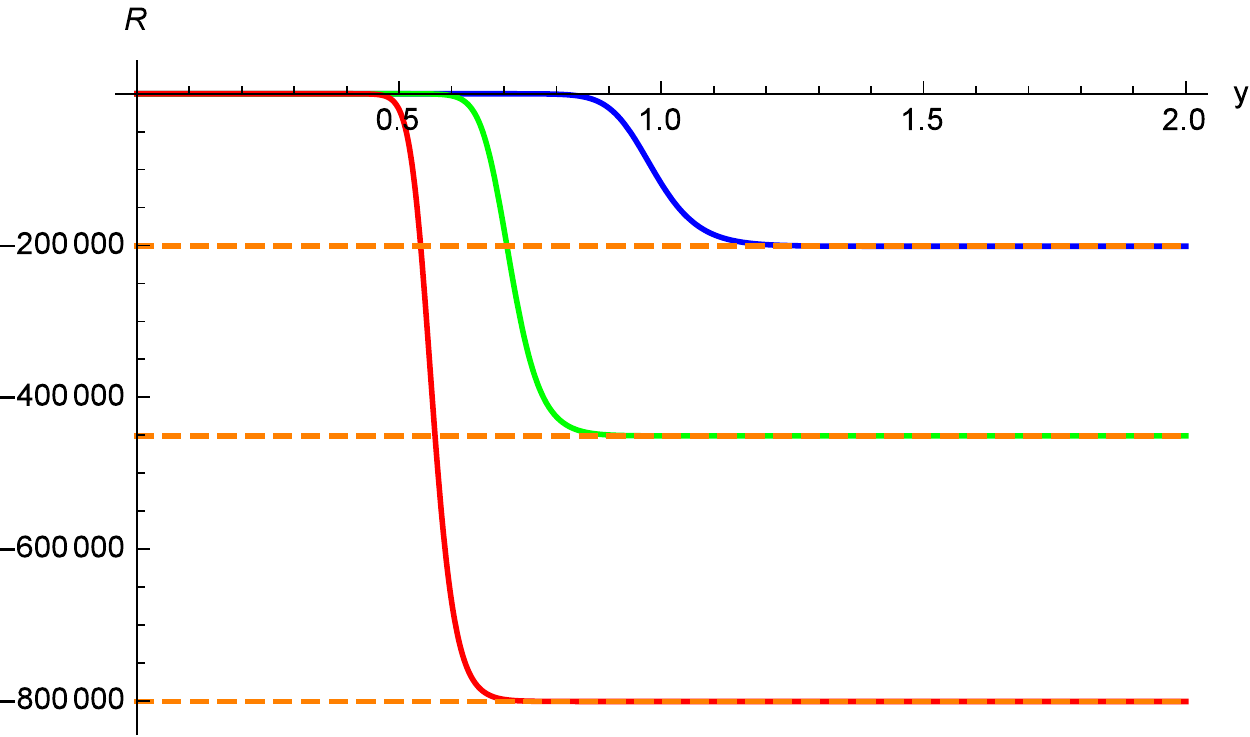, width=0.4\textwidth}}
\vspace*{4pt}
\caption{Ricci scalar $R \equiv g^{\mu\nu}R_{\mu\nu}(g)$ of EiBI gravity solution in the case $\epsilon<0$
with $\k^2=1$, $\e=-10^{-3}$ and $\b=-20$ (blue), $-30$ (green) and $-40$ (red). 
Dashed (orange) line represent the values in the central region($y \to \infty$) for each case.} 
\label{fig:Ricci}
\efig

\noindent{\it Energy.} At first sight, the $\e<0$ branch of the solution seems to have also a completely nice behavior in terms of energy distribution.
Indeed, looking at the transition region of the object (outside-inside), 
the effective energy density for the matter generated by its geometry ($\k^2\r_{eff}= - {G^t}_t$)
result finite everywhere and appears distributed on a thick shell around  $r\lessapprox |\b|$, while
the canonical energy density diverges at that point in the $\e>0$ case (see Fig. \ref{fig:EiBI_density} for a comparison). 
So, apparently, we have an interior region that looks like a ball with constant negative energy density, free of pathologies. 
However, the contributions of order $\k^2$ cannot be disregarded when approaching the center of the solution, as the full Einstein tensor ${G^t}_t$ component take the form
\be 
{G^t}_t=\tfrac{3}{2 |\e|\k^2}\left(\b^2+2\b\g +\k^2\right)-\tfrac{1}{4\g^2}e^{\left(\b +2\g \right) y} \ ,
\ee
and divergent contributions emerge in the $y\to \infty$ limit.

\begin{figure}[tbp]
\centerline{\psfig{file=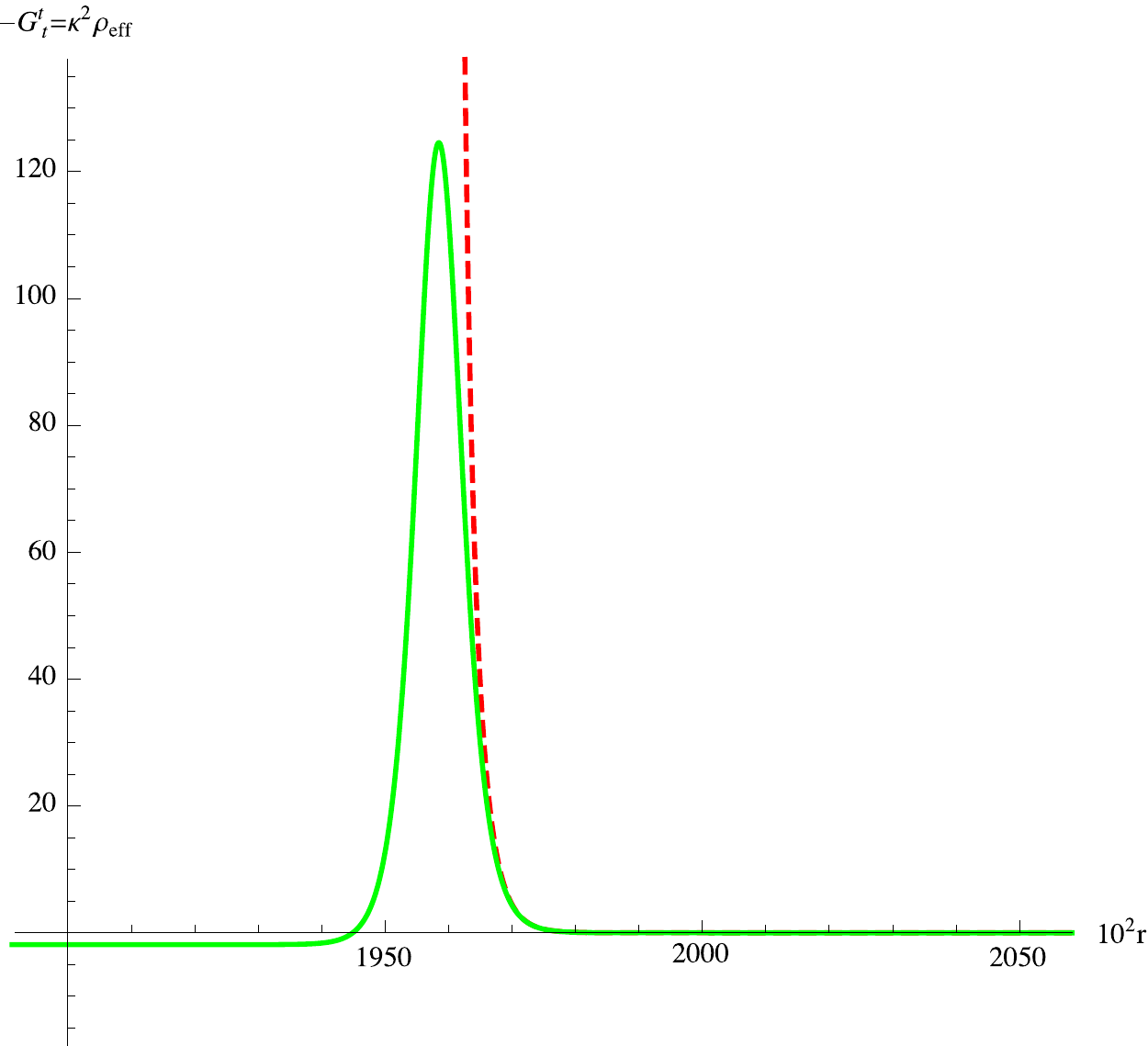,width=5.3cm, height=4cm}}
\vspace*{2pt}
\caption{Effective energy density $\k^2\r_{eff}\equiv-{G^t}_t$ of the EiBI solutions for $M=10$ (in units of $\k^2$) for the divergent $\e>0$ branch (dashed, red) and the well-behaved $\e<0$ branch (solid, green).  Note the location of the peak at a radius below the Schwarzschild radius, here at $10^2r=2000$.}
\label{fig:EiBI_density}
\efig

\medskip
\noindent{\it Curvature.}
The scalar curvature of our EiBI solution shows its own additional divergences.
In fact, the full expression that in the astrophysical approximation reads
\be\label{eq:Ricci_EiBI}
R=-\tfrac{5}{2 |\e|\k^2} (\beta ^2+\tfrac85\beta \g + \k^2)
+\tfrac{1}{2\g^2} e^{\left(\b +2\g \right) y}  \ ,
\ee
diverges as\footnote{The compact expressions of the Ricci scalar and the effective energy density in the regions of interest are obtained by approximating in the metric the hyperbolic functions by exponentials, which is extremely accurate in all relevant cases, as can be easily verified numerically.}
 $R\sim e^{(\k^2/|\b|)y}$,
which is much softer than the GR solution divergence, 
\be\label{eq:R_GR}
\lim_{y\to\infty} R_{GR}\approx \frac{\k^2}{\left(\beta ^2+2 \kappa ^2\right)^2} e^{ \left(2 \sqrt{\beta ^2+2 \kappa ^2}+\beta \right)y}
\sim e^{|\beta| y}\ .
\ee
\smallskip
We thus learn that, for astrophysical sources, while this divergence is strongly magnified in GR, it remain strongly suppressed  in the EiBI theory.

\bigskip
To get a grasp of the physical implications of the above divergences, it is instructive to consider the energy density 
from the point of view of the matter sector, {\it i.e.}, the stress-energy tensor of the scalar field. 
On the GR side we have $\rho_{GR}\equiv -{T^t}_t= Z/2=q^{yy}/2$, and thus the internal region of the solution diverges like $\rho_{GR} \sim e^{|\beta| y}$. 
This coincides with what is shown by the geometric sector in \eqref{eq:R_GR}, and is nothing but a reflection of the Einstein's equations structure.
On the EiBI side instead, we have $\rho_{EiBI}=L(X)/2$, which tends to $\rho_{EiBI}\approx  1/(|\epsilon|\kappa^2)$ in the internal region.
That is, the central region energy density is a positive constant that saturates the natural scale of the EiBI Lagrangian \eqref{eq:PZBId}. As depicted in Fig. \ref{fig:kink}, this scalar field energy density $\rho_{EiBI}$ presents a `kink-like' profile
completely different to the behavior of the $\e>0$ case, also depicted in the graph.

\begin{figure}[tpb]
\centerline{\psfig{file=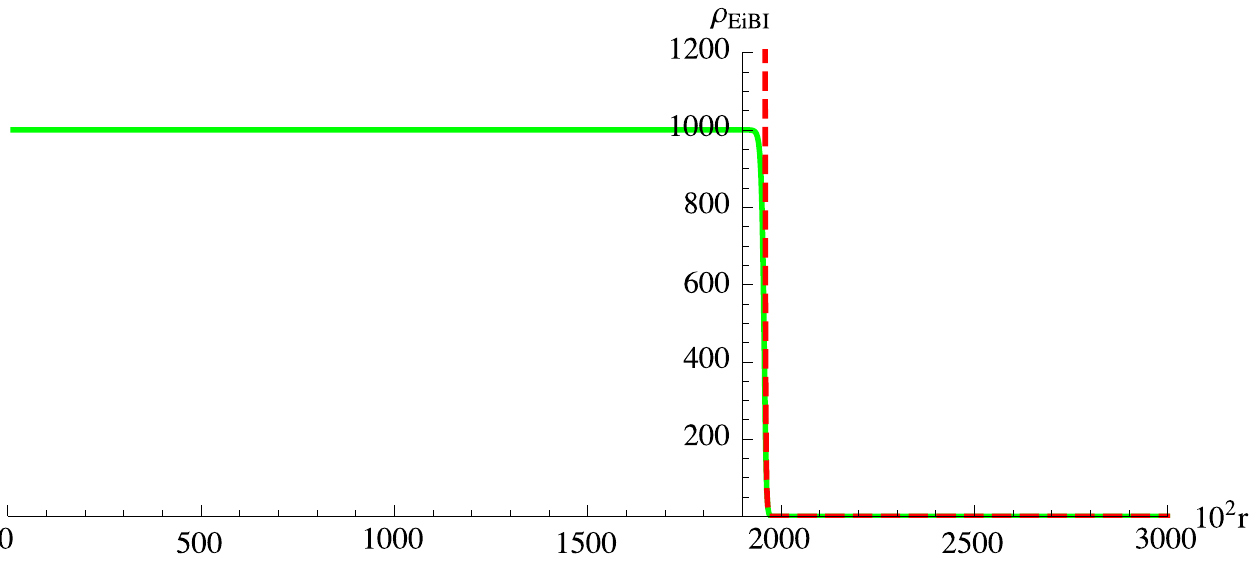,height=3.5cm,width=6cm}}
\vspace*{4pt}
\caption{Energy density $\rho_{EiBI}\equiv-{T^t}_t$ of the scalar field matter.
For $\e<0$ (solid, green), the transition from the vacuum value $\r_{EiBI}=0$ to the internal maximum $\r_{EiBI}=1/(|\e|\k^2)$ takes place below the corresponding Schwarzschild radius $r\lessapprox |\beta|$, here located at $10^2r=2000$. 
For $\e>0$ (dashed, red), the canonical energy density diverges at that point.}
\label{fig:kink}
\efig

The key result here resides in the fact that the interpretation of the (divergent) Einstein tensor as an observable 
effective stress-energy tensor provides a physical picture completely dissociated of the actual behavior of the matter field 
which shows, at least in terms of it energy density, a physically acceptable picture.
This points to the highly non trivial question of whether the Ricci scalar and Einstein tensor divergences 
carry any valuable physical content, which has already been risen in works dealing with other metric-affine spaces.\cite{Olmo:2016fuc}

\bigskip{\it Geodesics.}
The geodesic structure of the spacetime will give us the final ingredients to complete our picture of the solution.
As stated in Sec. \ref{sec:II}, in RBGs matter particles would follow metric geodesics.
For a spherically symmetric metric written as $ds^2=-C(x) dt^2+D(x)^{-1} dx^2+r^2(x)d\Omega^2$, the geodesic equation can be put in the form\cite{LecOlmo}
\be \label{eq:geoeqo}
\tfrac{C(x)}{D(x)} \left(\tfrac{dx}{d\l}\right)^2=  \E^2 -C(x) \left(\tfrac{\ell^2}{r^2(x)} -k \right),
\ee
with $k=1,0,-1\,$ for space-like, null and time-like geodesics, respectively.
For time-like geodesics, the conserved quantities $\E\equiv\sqrt{DC} dt/d\l$ and $\ell\equiv r^2(x)d\varphi/d\l$ 
are interpreted (due to staticity and spherical symmetry) as the total energy per unit mass and the angular momentum per unit mass\footnote{Around an axis normal to a plane that, without loss of generality,  can be taken to be $\th=\pi/2$.}, respectively.
In the case of null geodesics, it is the quotient $\ell/\E$ that can be identified with an apparent impact parameter from asymptotic infinity \cite{Chandra}.

The energy conservation condition for the line element \eqref{eq:BImapping} reads $\E=e^{\nu}dt/d\lambda$, while the radial null geodesics satisfy
\be
\(\tfrac{d(\E\lambda)}{dy}\)^{\,2}={e^\nu}\left({{e^\nu}W^{-4}+|\epsilon|\kappa^2}\right) \ .
\ee
In the case under study, the geometry of the asymptotically flat GR solution
must be recovered away from the region of localization of the scalar field.
Indeed, far from the center ($y\to 0$), we have $e^\nu\approx 1$ and $W=1/r\approx y$. Therefore $(dr/d\lambda)^2=1$
and $r(t)=r_0\pm t$, which represents light rays propagating at the speed of light ($c=1$), as expected.
Thus, it is only relevant to discuss the geodesics in the interior region.

In the limit towards the center  ($y\to \infty$), $(d(\E\l)/dy)^2\!\approx\! |\e|\k^2 e^{-|\b| y}$, which integrates to 
$\E\D \l=\mp 2|\b|^{-1}\!\sqrt{|\e|\k^2} e^{-|\b|y/2}$. 
This result implies a nonconventional behavior:
$dy/d\l$, the effective speed of the light rays inside the object, grows exponentially fast near the center (see Fig. \ref{fig:geodesics_EiBI}),
leading to a ray that travels from the surface of the object at $y_c$, through the center to its antipodal point,
in an absurdly short affine time, $\Delta \lambda \propto e^{-|\beta|y_c/2}$.
This behavior is paralleled by non-radial rays and massive particles (time-like geodesics),
whose geodesic equations in the interior region result degenerated with those of the radial null geodesics.

\begin{figure}[tpb]
\centerline{\psfig{file=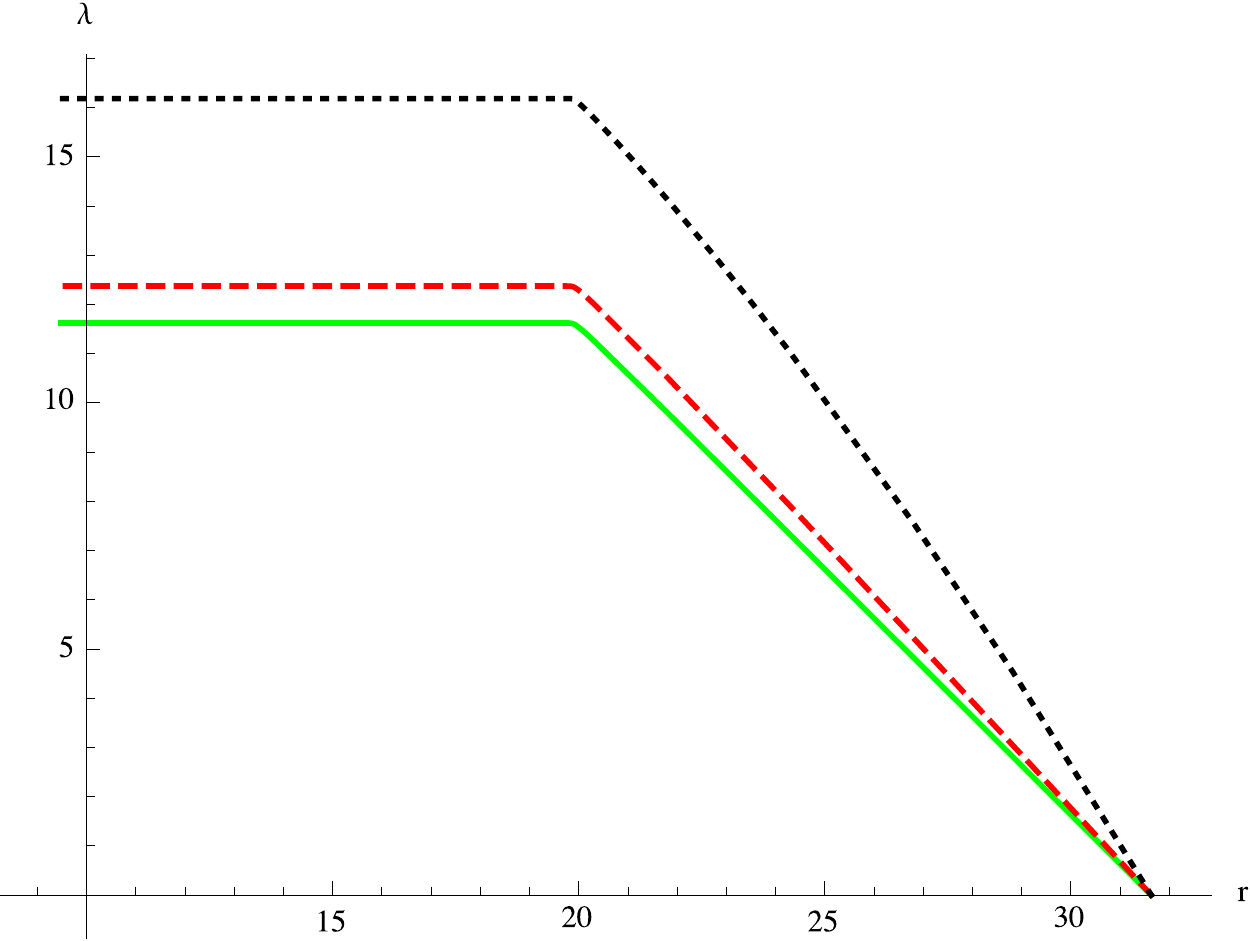,height=0.16\textheight,width=0.38\textwidth}}
\vspace*{4pt}
\caption{Ingoing geodesics for $\e<0$ case: radial null (solid, green), null with angular momentum (dashed, red), time-like with angular momentum (dotted black). After crossing the surface of the object ($\sim r\lessapprox 20$), light cones get dramatically modified allowing for an almost instantaneous transfer of particles and information between antipodal points of the compact object. This behavior can be interpreted as a wormhole but with Euclidean topology. }
\label{fig:geodesics_EiBI}
\efig

Contrary to what we observed in the $\e>0$ case, which presents an ill-defined  interior geometry as a result of the divergence of the energy density at the surface and the inversion of the metric signature (recall Figs. \ref{fig:EiBI_density} and \ref{fig:kink}),
the weird behavior of the $\e<0$ branch, namely, that information and particles can traverse the object almost instantaneously, 
even when the (proper) radial distance from the surface to the center is infinite (recall our computation \eqref{eq:properdistance}) has, nonetheless, a physically plausible interpretation.

If we would determine the physical proper radial distance $\Delta L$ (from the surface) to the center of the object 
operationally, that is, by assuming a constant speed of light $c$ and computing the proper time $\D\l$ for a signal traveling that journey,
we would have concluded that $\Delta L=c \D\l \to0$, as if the interior did not exist.
Thus, in practical terms, the $\e<0$ branch of the solution describes an object that functions like a kind of \emph{wormhole membrane}, which instantaneously transfers particles and information between antipodal points of its surface. 
This is in good agreement with results previously obtained in \cite{Afonso:2017aci}, where the same RBG system was studied without the mapping technology, but the emergence of wormhole structures was clearly indicated by a mix of numerical and approximative solutions of the field equations.

\section{Conclusions and perspectives} \label{sec:VI}

In the present work we have obtained and analyzed the features of exact scalar field solutions describing gravitating compact objects in the Eddington-inspired Born-Infeld gravity, a well motivated metric-affine extension of GR.
We focused on this model with a two-fold objective in mind.
First, to illustrate, by means of a particular case, common features shared by the members of a large class of extensions of GR in the metric-affine formalism that we call Ricci-based gravities.
These models, constructed out of scalars based on the Ricci tensor and the metric, support an Einstein-frame representation which allows to establish a correspondence between the RBG's and GR's field equations.
Our second goal was to show in detail how this correspondence, first presented in \cite{Afonso:2018hyj},
 can be exploited as an extremely efficient tool to generate solutions for the highly non-linear, and otherwise hardly solvable, field equations that arise in RBG theories.

Exact solutions for the EiBI gravity coupled to (non-canonical) scalar Lagrangians were generated using 
as seed the static, spherically symmetric solution obtained by Wyman for a free (canonical) scalar field in GR. 
Two branches where found, depending on the sign of $\e$, the single parameter of the model. 
Despite the fact that the line element of this solution turns out to be quite similar to the original one,
the physical consequences are profound.
For instance, while the radial proper distance to the center ($y\to\infty$), calculated with the Wyman solution
shortens exponentially fast as $l_{GR}\propto e^{-( \b+4\g) y}$,
for the $\e<0$ EiBI solution in the same approximation we have $l_{EiBI}=\sqrt{|\epsilon|}\kappa y$,
which puts the center at an infinite proper radial distance,
showing that the internal structure of these objects is radically different from their GR correspondent. 

For astrophysical configurations ($\vert\b\vert^2 \gg \k^2$), the objects found are strictly horizon-free but present, nonetheless, some distinguishing characteristics slightly below the would-be Schwarzschild radius of an equivalent black hole solution with the same mass. 
For instance, in the $\e>0$ case, the $y$ coordinate is not allowed to surpass a maximum value, $y_{max}=- \log[{\vert\e\vert \k^2}/{(2\k^2+\b^2)^2}]$, so this object could be described as a `rigid shell with no interior'.
If we agree that a physically consistent spacetime is one where neither information (null geodesics) nor physical observers (time-like geodesics) should disappear or emerge out of nowhere, then this branch of the solution is an ill-defined spacetime.
In fact, its structure avoids geodesics to be extended to arbitrarily large values of their affine parameter, constituting a geodesically incomplete spacetime.

The negative ($\e<0$) branch, instead, seems to describe a much more interesting object,
admitting a double interpretation depending on which definition is adopted to describe the energy density distribution.
On the one hand, looking at the canonical stress-energy tensor of the scalar matter one identifies a localized object: 
a vanishing density outside a well defined radius ($r\lessapprox 2M$),
and then an interior constant maximum positive density reached after a quick transition.
On the other hand, one could interpret the Einstein tensor as an effective stress-energy tensor. 
Under this approach the object looks like a thin shell supported by a negative interior pressure.
From an observational point of view, this new kind of object behaves as a `wormhole membrane', 
transferring particles and light from any point on the surface to its antipodal point in a vanishing fraction of proper time.

\smallskip Some remarks on more general aspects are in order.
When applied to electric fields coupled to RBGs, the mapping technique 
also produced solutions interpreted as wormholes and objects without interior (see e.g. \cite{Bambi:2015zch}).
However, the energy density in electrovacuum solutions gets concentrated mostly around the center.
While this distribution gives rise to distinctive properties  on the background geometry of the interior regions\cite{Menchon:2017qed}, the structure of external horizons is substantially preserved.
Consequently, no effects could be observed on astrophysical scales.

The case of the scalar field objects seems to be fundamentally different,
as the energy density distribution results appreciable modified already near the would-be Schwarzschild radius.
This implies that non-GR effects already occur at macroscopic, astrophysical scales.
Thus, even when no observable distinction with respect to a Schwarzschild black hole could be identified 
 in, for instance, the orbital motions around these objects (recall that GR$+\L$ is exactly recovered in vacuum regions), 
 their shadows would present distinguishable signals due to the lack of event horizons \cite{Cunha:2015yba}.
This puts this kind of objects on the list of candidates to black hole mimickers.

The lesson we learn from this is that modified gravity theories like RBGs,
whose gravitational dynamics depend on the \emph{local} stress-energy densities,
may lead to departures from GR even at observationally relevant scales 
and not only at Planck scale, as usually assumed.

\smallskip The results obtained have motivated us to extend our explorations in two main directions.
First, to develop a systematic search for new compact objects in other Ricci-based gravities (for instance, $f(R,Q)$ theories are under study). 
Second, to carry out the stability and perturbative analysis of the kind of solutions here discussed,
looking for a characterization of their distinctive features, 
which is mandatory in order to gain the ability to identify their possible signals in astrophysical observational data, 
for instance, against gravitational wave data as proposed in \cite{Cardoso:2016oxy,Konoplya:2019nzp}.
Progress along these lines will be reported soon. 

\section*{Acknowledgments}
The author would like to thank Prof. Luis Carlos Bassalo Crispino for his kind invitation to contribute to the ``V Amazonian Symposium on Physics'' (V ASP 2019) by presenting a talk, on which this article is based, and to the Department of Physics of Universidade Federal do Par\'a (Brazil) for the hospitality and support during the stay.
The author is partially supported by Federal University of Campina Grande, Brazil and this work was partially funded  by the Spanish project SEJI/2017/042 (Generalitat Valenciana).  This article is based upon work from COST Action CA15117, supported by COST (European Cooperation in Science and Technology).

\end{document}